\documentclass[aps,prl,preprint,superscriptaddress]{revtex4}

\usepackage[breaklinks=true,colorlinks,citecolor=blue,linkcolor=blue,urlcolor=blue]{hyperref}
\usepackage{amssymb,amsmath,amsfonts}
\usepackage{subfigure, latexsym,multirow}
\usepackage[notcite,color,final]{showkeys}
\usepackage{epsfig,graphicx,subfigure,dcolumn,bm}
\usepackage[usenames ,dvipsnames]{xcolor}
\usepackage{slashed}
\usepackage{ulem}
\begin{document}
\title{Relativistic horizon of interacting Weyl fermions \\ in condensed matter systems}
\author{Wei-Chi Chiu $^*$\footnote[0]{*These authors contributed equally to this work.} }
\affiliation{Department of Physics, Northeastern University, Boston, MA 02115, USA}

\author{Guoqing Chang $^*$,$^{\dag}$\footnote[0]{$^{\dag}$Corresponding author (email): guoqing.chang@ntu.edu.sg }}
\affiliation{Division of Physics and Applied Physics, School of Physical and Mathematical Sciences, Nanyang Technological University, Singapore, Singapore}

\author{Gennevieve Macam}
\affiliation{Department of Physics, National Sun Yat-sen University, Kaohsiung 80424, Taiwan}

\author{Ilya~Belopolski}
\affiliation {Laboratory for Topological Quantum Matter and Advanced Spectroscopy (B7), Department of Physics, Princeton University, Princeton, New Jersey 08544, USA}
\affiliation {RIKEN Center for Emergent Matter Science (CEMS), Wako, Saitama 351-0198, Japan}

\author{Shin-Ming Huang}
\affiliation{Department of Physics, National Sun Yat-sen University, Kaohsiung 80424, Taiwan}

\author{Robert Markiewicz}
\affiliation{Department of Physics, Northeastern University, Boston, MA 02115, USA}

\author{Jia-Xin Yin}
\affiliation {Laboratory for Topological Quantum Matter and Advanced Spectroscopy (B7), Department of Physics, Princeton University, Princeton, New Jersey 08544, USA}

\author{Zi-jia Cheng}
\affiliation {Laboratory for Topological Quantum Matter and Advanced Spectroscopy (B7), Department of Physics, Princeton University, Princeton, New Jersey 08544, USA}

\author{Chi-Cheng Lee}
\affiliation {Department of Physics, Tamkang University, Tamsui, New Taipei 251301, Taiwan}

\author{Tay-Rong Chang}
\affiliation {Department of Physics, National Cheng Kung University, Tainan, Taiwan}
\affiliation {Center for Quantum Frontiers of Research and Technology (QFort), Tainan, Taiwan}
\affiliation {Physics Division, National Center for Theoretical Sciences, Taipei 10617, Taiwan}

\author{Feng-Chuan Chuang}
\affiliation {Department of Physics, National Sun Yat-sen University, Kaohsiung 80424, Taiwan}
\affiliation {Physics Division, National Center for Theoretical Sciences, Taipei 10617, Taiwan}
\affiliation {Department of Physics, National Tsing Hua University, Hsinchu 30013, Taiwan}

\author{Su-Yang Xu}
\affiliation {Department of Chemistry and Chemical Biology, Harvard University, Cambridge, MA, USA.}

\author{Hsin Lin}
\affiliation{Institute of Physics, Academia Sinica, Taipei 11529, Taiwan}

\author{M.~Zahid~Hasan}\affiliation {Laboratory for Topological Quantum Matter and Advanced Spectroscopy (B7), Department of Physics, Princeton University, Princeton, New Jersey 08544, USA}
\affiliation{Lawrence Berkeley National Laboratory, Berkeley, California 94720, USA}

\author{Arun Bansil}
\affiliation{Department of Physics, Northeastern University, Boston, MA 02115, USA}

\date{\today}

\begin{abstract}

The intersections of topology, geometry and strong correlations offer many opportunities for exotic quantum phases to emerge in condensed matter systems. Weyl fermions, in particular, provide an ideal platform for exploring the dynamical instabilities of single-particle physics under interactions. Despite its fundamental role in relativistic field theory, the concept of causality and the associated spacetime light cone and event horizon has not been considered in connection with interacting Weyl fermionic excitations in quantum matter. Here, by using charge-density wave (CDW) as an example, we unveil the behavior of interacting Weyl fermions and show that a Weyl fermion in a system can open a band gap by interacting only with other Weyl fermions that lie within its energy-momentum dispersion cone. In this sense, causal connections or interactions are only possible within overlapping dispersion cones and each dispersion cone thus constitutes a solid-state analogue of the more conventional `event horizon' of high-energy physics. Our study provides a universal framework for considering interacting relativistic quasiparticles in condensed matter by separating them into energy-like and momentum-like relationships in analogy with the time-like and space-like events in high-energy physics. Finally, we consider two different candidate materials for hosting the Weyl CDW phase: (TaSe$_4$)$_2$I and Mo$_3$Al$_2$C. Our study greatly enriches the phenomenology and unveils new connections between condensed matter and high-energy physics.

\end{abstract}

\maketitle

The intriguing connections between high energy and condensed matter physics have led to a deeper understanding of quantum matter \cite{Hasan2010, Qi2011, Bansil2016,  Weyl2018, Interacting, Cao_graphene, Andrei:2020tp, Yin:2021vo,  Su-Yang:2015wz,LV2015, MIT_Weyl, Franz:2018uf}. One such connection manifests itself in topological materials where relativistic particles can emerge as quasiparticles. A familiar example is the Weyl fermion, a massless spin-1/2 particle proposed in 1929 ~\cite{Weyl:1929uo}, which has been realized in many condensed matter systems~\cite{Su-Yang:2015wz, LV2015,MIT_Weyl}. Weyl fermions have also attracted recent attention due to their unique quantum responses, such as the quantized circular photogalvanic effect ~\cite{Hasan:2021uy, Nenno:2020wg, Ding2021,Moore2, RhSi,RhSi_pho,CoSi_pho}. Another frontier concerns strongly interacting systems which host unusual effects driven the interplay of correlations, topology and geometry ~\cite{Hasan2010, Qi2011, Bansil2016, Yin:2021vo, Andrei:2020tp, Cao_graphene, Interacting}. Here, correlated Weyl semimetals provide a perfect platform for exploring interaction effects on single-particle physics. The separation of the individual Weyl nodes with opposite topological charges in momentum space makes it impossible to hybridize the nodes and produce a fully gapped insulating state without violating symmetries. In correlated systems, interactions might make it possible to open a global gap in the system with Weyl fermions. For instance, it has been reported that a Weyl semimetal can be gapped out into an axion insulator by the charge-density-wave (CDW) pairing interaction ~\cite{Gooth:2019wf,Shi:2021aa,PhysRevB.102.115159, PhysRevResearch.2.042010}, although the general mechanism of this metal-insulator transition remains elusive. Even though Weyl fermions are rooted in quantum field theory, how relativistic physics \cite{Enstien1905} enters the interaction dynamics of Weyl semimetals remains unexplored.

\begin{figure*}[th]
\centering
\includegraphics[width=0.8\textwidth]{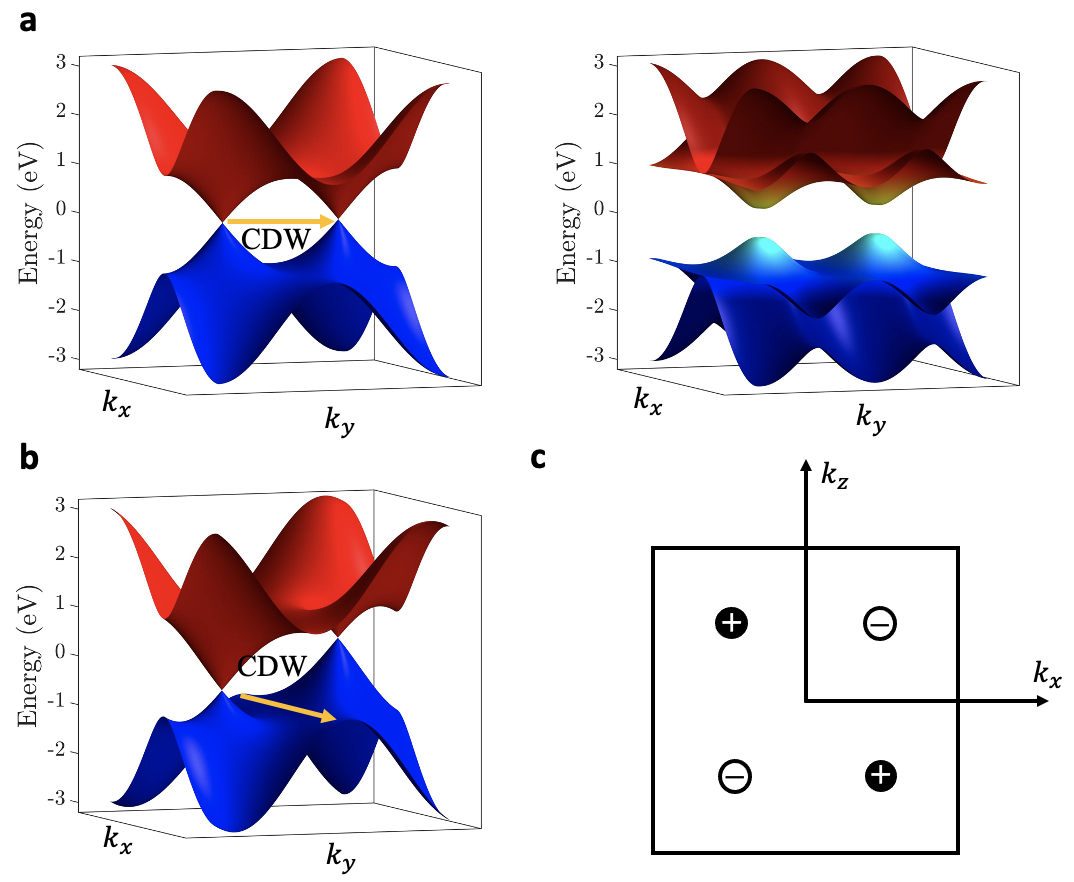}
\caption{\textbf{Schematic illustration of a Weyl semimetal with a CDW.} \textbf{a} Left panel: Special case of a pair of Weyl nodes of opposite chirality lying at the same energy, which are nested perfectly by the CDW Q-vector (yellow arrow). Right panel: Doubled energy bands and the resulting insulator driven by the CDW pairing-interaction. \textbf{b} The case of two Weyl nodes with an energy difference where the CDW Q-vector is not the same as the separation of two Weyl nodes. \textbf{c} Positions of the four Weyl nodes in our model and their chiralities.}
\label{fig:Weyl_1}
\end{figure*}

With this motivation, we discuss a topological phase transition mechanism for a CDW correlated Weyl semimetal system. In the existing literature, only the special case where the CDW wavevector is equal to the separation of the Weyl nodes in the momentum space has been considered (Fig.~\ref{fig:Weyl_1}a)~\cite{PhysRevB.102.115159,PhysRevResearch.2.042010}. How the possible energy difference between the Weyl nodes could affect the pairing interaction is not clear. In this work, we discuss the general case of a CDW in a three-dimensional Weyl semimetal (Fig.~\ref{fig:Weyl_1}b), which is much more representative of real materials~\cite{Gooth:2019wf,Shi:2021aa}, where the energy difference between the Weyl nodes and the momentum mismatch between the CDW Q-vector, along with the separation between the correlated Weyl nodes are all taken into account.

\textbf{Model and Methodology}

We start from a tight-binding model with intact time-reversal symmetry but broken inversion symmetry:
\begin{equation}
\mathcal{H}(k)=A\sin k_{x}\sin k_{z}\sigma_{0} +\left( \cos k_{x}-\cos k_{1}\right)  \sigma_{x} +\sin k_{y}\sigma_{y} +\left( 1+\cos k_{z}-\cos k_{y}\right)  \sigma_{z}, 
\end{equation}
where $A$ and $k_{1}$ are two constants with $k_{1}\neq \pi$, and  the $\sigma$ are the Pauli matrices. The positions of four Weyl nodes with $\mp1$ chirality are at $\bm{k}_{W}=\pm \left( k_{1,}\  0,\  \pm \pi/2 \right) $ as shown in Fig.~\ref{fig:Weyl_1}d with black (white) dots representing positive (negative) chirality.  The Weyl nodes of different chirality are at energies  $E_{W}=\pm A \sin k_{1}$.

To discuss the dynamics of interacting Weyl fermions, we consider the CDW instability as a quasi-one-dimensional Peierls instability such that there is only one unidirectional CDW Q-vector ($\bm{Q}_{CDW}$). For simplicity, but without losing generality of our theory, in all of our discussions on the tight-binding model, we fix the $\bm{Q}_{CDW}$ and vary the Weyl nodes separation instead. We choose $\bm{Q}_{CDW}=(\pi, 0, 0)$ as a representative, which reflects the Peierls dimerization in a double supercell along the $x$-direction. (See Methods section).

\begin{figure*}[th]
\centering
\includegraphics[width=0.8\textwidth]{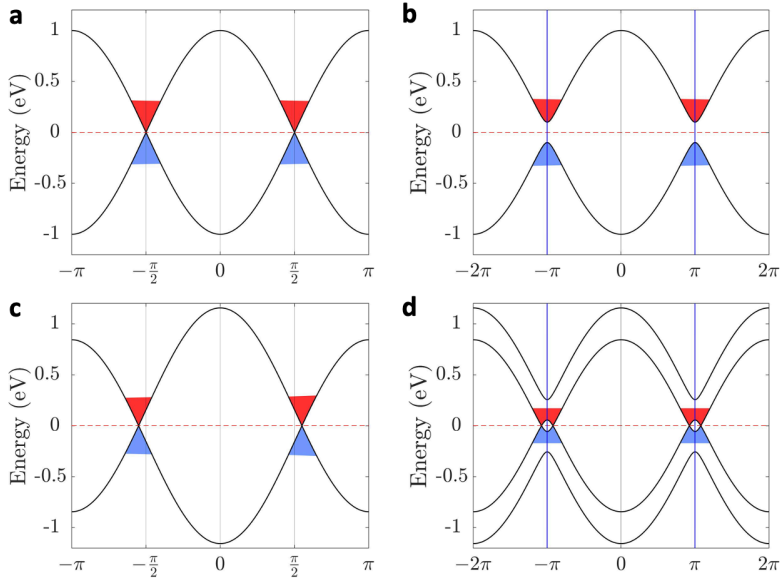}
\caption{\textbf{Weyl-CDW interacting when the nodes lie at the same energy.}  \textbf{(a-b)}. Band structure as a function of $k_x$. $A=0$, $k_1=\pi/2$, $k_y=0$, and $k_z=\pi/2$. \textbf{a}. Without CDW, four Weyl nodes are at the $\pm \left( \pi/2, 0, \pm \pi/2\right)$ with no energy difference. The red/blue color blocks represent the conduction/valence bands of Weyl cones. \textbf{b}. With CDW, CDW Q-vector is along  $\left( \pi, 0, 0 \right)$ and the CDW strength $\delta=0.05$. Weyl nodes with opposite chirality are nested exactly to the same point in energy-momentum space. There is a global gap opened by CDW between conduction and valence bands. The blue lines represent the boundary of the reduced BZ. \textbf{(c-d)}. Bands structure along the $k_x$ with $A=0$, $k_1=1.1\pi/2$, $k_y=0$, and $k_z=\pi/2$. \textbf{c}. Without CDW, four Weyl nodes are at the $\pm \left( 1.1\pi/2, 0, \pm \pi/2\right)$ \textbf{d}. With CDW, CDW Q-vector is along  $\left( \pi, 0, 0 \right)$ and the CDW strength $\delta=0.05$. The Weyl nodes cannot be gapped and the system remain in semimetal phase.}
\label{fig:Weyl_2}
\end{figure*}

We first consider the case $A=0$, where four Weyl nodes at $\pm \left( \pi/2, 0, \pm \pi/2\right)$  are at the same energy ($E_{W_1}=E_{W_2}$) (Fig.~\ref{fig:Weyl_2}a). After including the Peierls' dimerization with the CDW strength $\delta=0.05$, each pair of Weyl nodes along the $k_x$-direction are folded to the same point ($\bm{Q}_{CDW}=\bm{k}_{W_1}-\bm{k}_{W_2}=(\pi, 0, 0)$). As a result, the Weyl nodes annihilate with each other and open a global gap between the conduction and valence bands in the reduced BZ (Fig.~\ref{fig:Weyl_2}b). This is consistent with previous work, where CDW interactions can gap out Weyl semimetals ~\cite{PhysRevB.102.115159,PhysRevResearch.2.042010}. Now, we consider another interacting scenario where CDW Q-vector is different from the separation of Weyl nodes ($\bm{Q}_{CDW} \neq \bm{k}_{W_1}-\bm{k}_{W_2}$) (Fig.~\ref{fig:Weyl_2}c). By choosing $k_1=1.1\pi/2$, the separation vector for the pair of Weyl nodes along the $k_x$-direction is $(1.1\pi, 0, 0)$. After the inclusion of a CDW interaction with $\delta=0.05$, the paired Weyl nodes are folded at different locations in momentum space. Weyl nodes do not annihilate with each other and the system remains in the semimetal phase (Fig.~\ref{fig:Weyl_2}d). Our calculations show that the relation between Q-vector and the separation of Weyl nodes played a key role in determining the topological phase transition of correlated Weyl semimetals.

\begin{figure*}[th]
\centering
\includegraphics[width=1\textwidth]{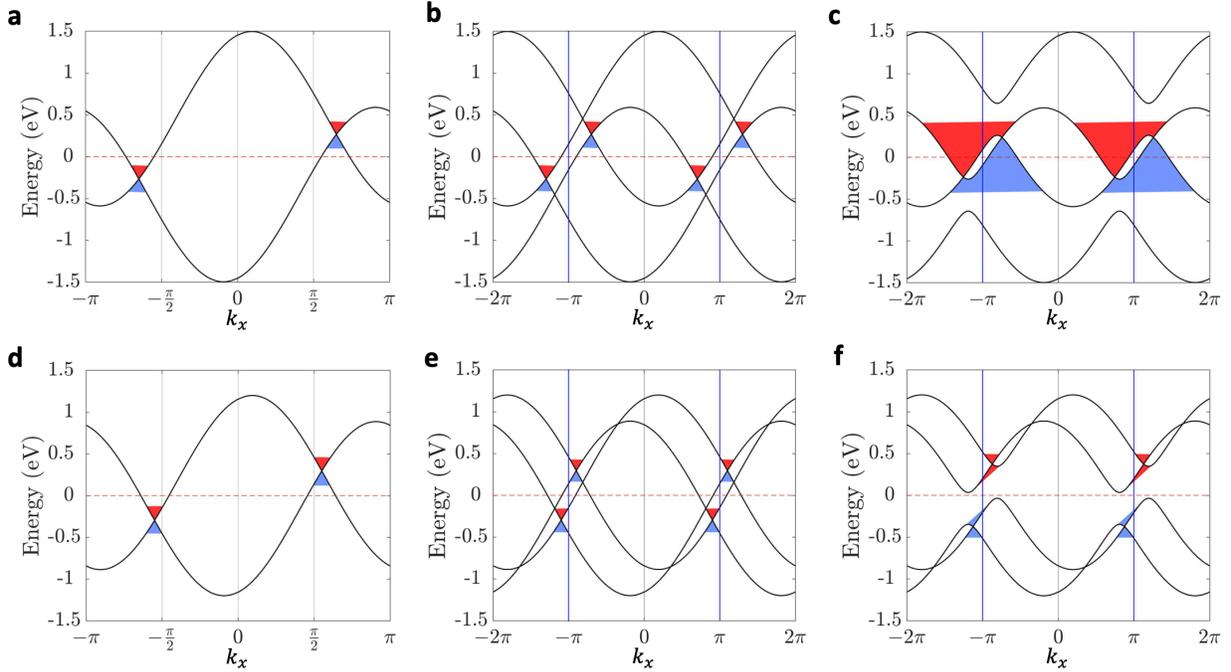}
\caption{\textbf{Weyl-CDW interacting when the nodes lie at different energy.} \textbf{(a-c)}. Bands structure as a function of $k_x$ with $A=0.3$, $k_1=1.3\pi/2$, $k_y=0$, and $k_z=\pi/2$. \textbf{a}. Without CDW, four Weyl nodes are at the $\pm \left( 1.3\pi/2, 0, \pm \pi/2\right)$ with an energy difference around 0.6 eV. \textbf{b}. Without CDW ($\delta=0$), the folded bands in the double supercell BZ along x-direction. Weyl nodes with opposite chirality are nested out of each other's cone. \textbf{c}. With CDW, CDW Q-vector is along  $\left( \pi, 0, 0 \right)$ and the CDW strength $\delta=0.1$. The Weyl nodes cannot be gapped and the system remain in semimetal phase. \textbf{(d-f)}. Bands structure along the $k_x$ with $A=0.3$, $k_1=1.1\pi/2$, $k_y=0$, and $k_z=\pi/2$. \textbf{d}. Without CDW ($\delta=0$), four Weyl nodes are at the $\pm \left( 1.1\pi/2, 0, \pm \pi/2\right)$. \textbf{e}. Without CDW, the folded bands in the double supercell BZ along x-direction. Weyl nodes with opposite chirality are nested into each other's cone. \textbf{f}. With CDW, CDW Q-vector is along  $\left( \pi, 0, 0 \right)$ and the CDW strength $\delta=0.1$ (See SM for the $\delta=0.05$ case). A global gap opened by the CDW around the Fermi energy, and the whole system undergoes a topological phase transition.}
\label{fig:Weyl_3}
\end{figure*}

We then consider the scenario when the two interacting Weyl nodes with an energy difference ($E_{W_1} \neq E_{W_2}$). By choosing $A=0.3$ and $k_1=1.3\pi/2$, the four Weyl nodes are at the $\pm \left( 1.3\pi/2, 0, \pm \pi/2\right)$ with an energy difference around 0.6 eV (Fig.~\ref{fig:Weyl_3}a). To understand how the Weyl nodes are folded in the reduced BZ, the folded band structure for the double supercell along the $x$-direction when CDW interaction strength $\delta$  set as 0 is plotted in Fig.~\ref{fig:Weyl_3}b. The Weyl nodes are folded to the outside the dispersion-cone of each other. After the inclusion of nonzero CDW interaction strength $\delta$, we find the Weyl nodes do not annihilate with each other and the system remains in the metallic phase (Fig.~\ref{fig:Weyl_3}c). Here, we choose the $\delta=0.1$ for presentation reason. To figure out the condition on whether two Weyl nodes can annihilate with each other in the case  when $E_{W_1} \neq E_{W_2}$ and $\bm{Q}_{CDW} \neq \bm{k}_{W_1}-\bm{k}_{W_2}$, we change the location of Weyl nodes to $\pm \left( 1.1\pi/2, 0, \pm \pi/2\right)$ (Fig.~\ref{fig:Weyl_3}d). In this case, the Weyl nodes in the folded BZ are within the dispersion-cone of each other (Fig.~\ref{fig:Weyl_3}e). Surprisingly, after the inclusion of a nonzero CDW interaction strength $\delta$, a global gap between the conduction and valence bands in the system has been opened (Fig.~\ref{fig:Weyl_3}f).


\textbf{Discussion}

\begin{figure*}[th]
\centering
\includegraphics[width=1\textwidth]{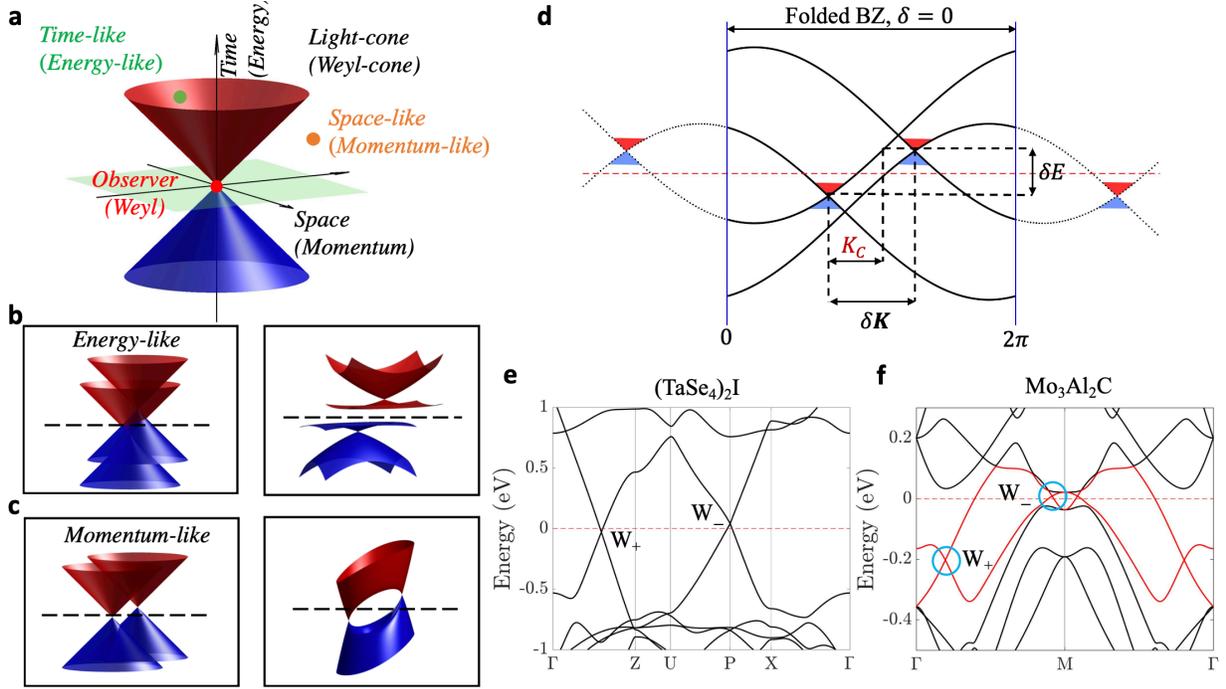}
\caption{\textbf{Weyl-CDW in analog with relativistic physics.} \textbf{a} In analogy with the theory of relativity, we define the region in energy-momentum space within/outside the Weyl-cone as energy-like/momentum-like region.  \textbf{b} In the left panel, two Weyl nodes are energy-like in reduced BZ. In the right panel, with including interaction, the system undergoes a semimetal-insulator phase transition. \textbf{c}  In the left panel, two Weyl nodes are momentum-like in reduced BZ. In the right panel, with including interaction, the system remains in a semimetal phase. \textbf{d} Schematic picture of critical length $K_C$ in the reduced BZ. \textbf{e} band structure of (TaSe$_4$)$_2$I from tight-binding model. A pair of Weyl nodes cross the Fermi energy. \textbf{e} DFT band structure of Mo$_3$Al$_2$C. The red curves represent the two bands with a pair of  Weyl nodes (highlighted in the blue circles) cross the Fermi level along the $\Gamma$-M direction.  }
\label{fig:Weyl_4}
\end{figure*}

We observe that whether the Weyl-CDW pairing interaction will drive the topological semimetal-insulator phase transition depends on the relative location of Weyl nodes in the reduced BZ. In analogy with relativistic physics, we define the region in energy-momentum space within (outside) the Weyl-cone as the energy-like (momentum-like) region (Fig.~\ref{fig:Weyl_4}a). Then we can define two classes as being energy-like or momentum-like, the left panels of Fig.~\ref{fig:Weyl_4}b and \ref{fig:Weyl_4}c, expressed as 
\begin{equation} 
\label{Weyl_rel}
\begin{cases} (\delta E/V_{F})^{2}-(\delta \bm{K})^{2}<0:\ \text{Momentum-like} \\ (\delta E/V_{F})^{2}-(\delta \bm{K})^{2}>0:\ \text{Energy-like} \end{cases}, 
\end{equation}
where $\delta \bm{K}=\bm{k'}_{W_1}-\bm{k'}_{W_2}$, and $\bm{k'}_{W_1}$ and $\bm{k'}_{W_2}$ are the new Weyl nodes positions in the reduced BZ, $\delta E=E'_{W_1}-E'_{W_2}$ is the energy difference of two Weyl nodes with the CDW interaction, and $V_{F}$ is the Fermi velocity of the Weyl cone. Here we used the sign convention of the Minkowski metric $\eta_{\mu \nu } =diag(1,\  -1,\  -1,\  -1).$ For simplicity in discussion, we assumed that two Weyl fermions have the same Fermi velocity, and the Fermi velocity of each Weyl cone is isotropic.

Based on this classification, we find that the Weyl semimetal can become an insulator only when the Weyl nodes around the Fermi level with CDW interaction are energy-like (the right panel of Fig.~\ref{fig:Weyl_4}b). Note that, in this scenario, the Weyl nodes may not be gapped, but the whole system has a semimetal to insulator transition. In contrast, if two Weyl nodes are momentum-like, the Weyl system remains in the semimetal phase after the CDW phase transition. Hence, to examine whether a system can undergo the metal to insulator phase transition is equivalent to examining whether the Weyl nodes with opposite charges are energy-like in the CDW phase. 

Here, we give a brief discussion on the relation between the theory of relativity and our Weyl-CDW physical pictures. In Einstein's theory of relativity, space and time can not be described independently and they are connected by the speed of light. The causality means that a cause cannot have causal connection (effects) to an observer if it is not in the light cone of the observer. That is only when two events are time-like that they can be causally related. Accordingly, a horizon is a boundary in spacetime  beyond which events cannot affect an observer. Similarly, for a system with Weyl fermions, the duality of energy and momentum should be seen as a whole with their energy and momentum connected by its Fermi velocity. In contrast to the spacetime picture, in the energy-momentum space, only when two Weyl nodes are energy-like can they have correlation (causal connection) and make the whole system undergo a phase transition. The boundary of the loss of their correlation is the Weyl-cone which is a horizon in the energy-momentum space.

For more practical purposes, we further simplify Eq.~\eqref{Weyl_rel} in the case when the CDW interaction can be treated as a perturbation. We can assume that the energy and momentum of Weyl points only acquire a small correction from CDW interaction in its reduced BZ. We define a critical length in the energy-momentum space based on the energy difference of Weyl points and their Fermi velocity as $K_C=|{(E_{W_1}-E_{W_2})/V_{F}} |$. Hence, we can have a quick estimation of whether there is a metal-insulator transition in the Weyl-CDW system by simply comparing the critical length $K_C$ and the length of momentum separation after band folding without CDW ($\delta=0$)(Fig.~\ref{fig:Weyl_4}d). Note that, when the momentum separation is close to $K_C$, the CDW interaction $\delta$ may push the two Weyl quasiparticle toward each other and cross the quantum critical point, so detailed and careful CDW simulations are still needed.


Here, we emphasize that the causal structure we discussed is universal and not depending on the choice of model and the choice of $\bm{Q}_{CDW}$. We have done a series of tests based on different model and different $\bm{Q}_{CDW}$, and the causal structure of the interacting Weyl fermions are the same as what we concluded.

\textbf{Application to Real Materials}

Because Weyl fermions are quite normal in inversion symmetry breaking systems \cite{SAtran,KramersWeyl}, our theory can be widely applied to the big class of noncentrosymmetric CDW materials.
As an application of our arguments, we take the (TaSe$_4$)$_2$I as one example. (TaSe$_4$)$_2$I is a Weyl semimetal at room temperature and it turns into an incommensurate CDW phase with $\bm{Q}_{CDW} =(0.027 (\frac{2\pi}{a}), 0.027(\frac{2\pi}{a}), 0.012(\frac{2\pi}{c})) $ when the temperature is lowered below to $263$K~\cite{Shi:2021aa,Gooth:2019wf}. The Weyl nodes of (TaSe$_4$)$_2$I without SOC is shown in Fig.~\ref{fig:Weyl_4}e.  The energy difference of the two Weyl is $\delta E=0.068$\,eV with fermi velocity $V_F\sim3.47$\,eV$\cdot$\r{A}. Thus $K_C$ is estimated to be around $0.02\, \rm \AA^{-1}$. We take the approximation on the CDW supercell as the nearest commensurate supercell. Based on the folded band structure in this $37\sqrt{2}\times37\sqrt{2}\times 83$ commensurate supercell, we find the momentum difference between these two nodes is around $0.009\, \rm \AA^{-1}$, much smaller than $K_C$. Therefore, with the inclusion of CDW, the pair of Weyl nodes in (TaSe$_4$)$_2$I system are energy-like and the system will become an insulator. This is consistent with experimental measurements~\cite{Shi:2021aa}.

Our arguments could also be a possible explanation to reconcile the conflict on whether there is a CDW in the Mo$_3$Al$_2$C~\cite{PhysRevB.84.212501,PhysRevB.85.052501}. It has been argued that the sudden change in the electronic density of states is due to the Fermi surface nesting along the CDW nesting vector along with the $(1, 1, 1)$ direction~\cite{PhysRevB.84.212501}. However, it is also reported that there is no sign of the semimetal to insulator transition in the Mo$_3$Al$_2$C~\cite{PhysRevB.85.052501}. Here, we examine the band structure of the Mo$_3$Al$_2$C without including SOC. We find a pair of Weyl nodes along the $(1, 1, 0)$  ($\Gamma$-M) direction, on the bands cross the Fermi energy (Fig.~\ref{fig:Weyl_4}f). We find that the separation of folded Weyl nodes ($ \thicksim 0.23\, \rm \AA^{-1}$) are larger than $K_C  \thicksim 0.17\, \rm \AA^{-1}$, Mo$_3$Al$_2$C should be momentum-like with CDW interaction. Our argument indicates that the CDW in Mo$_3$Al$_2$C leads to the partial Fermi surface gap but no metal-insulator transition, which provides a possible explanation to reconcile the conflict in experiments.

\textbf{Conclusion}

We discuss phase transitions in a correlated Weyl semimetal by considering the dynamical instability of a CDW. In analogy with the spacetime light cone and the related causality-driven event horizon in relativistic physics, we unveil the causal structure of the energy-momentum space in the condensed matter context, which consists of energy-like and momentum-like regions. Our analysis reveals that the Weyl-CDW system can realize a metal-insulator transition only when a pair of interacting Weyl nodes are energy-like, otherwise they are forbidden to interact to produce a band gap. This physical picture of the Weyl-CDW system can also be directly applied to other topological systems with Weyl fermions. For example, Weyl physics can be simulated in a 3D optical lattice~\cite{PhysRevLett.114.225301}, where the CDW effect could be produced by introducing a period-2 superlattice (dimerization) using two additional orthogonal optical waves at double the in-plane wavelengths~\cite{PhysRevLett.111.185301}. In our case, the CDW is generated via spontaneous breaking of translational symmetry, which leads to the pairing interaction between the Weyl fermions. We emphasize, however, that in view of the universality of Weyl physics, our formalism will apply more generally to other pairing interaction that break the symmetries of an interacting system with relativistic quasiparticles in the energy-momentum space. Our study for the first time shows how the key concepts of causality and the associated event horizon in spacetime in relativistic physics can be carried over into the field of topological materials, and thus unveils new connections between condensed matter and high-energy physics.
 
\clearpage
\textbf{Competing interests}

The authors declare no competing interests.

\textbf{Acknowledgments}

G.C.  acknowledges the support of the National Research Foundation, Singapore under its Fellowship Award (NRF-NRFF13-2021-0010) and the Nanyang Assistant Professorship grant from Nanyang Technological University. The work at Northeastern University was supported by the Air Force Office of Scientific Research under award number FA9550-20-1-0322, and it benefited from the computational resources of Northeastern University's Advanced Scientific Computation Center (ASCC) and the Discovery Cluster. Work at Princeton University was supported by the Gordon and Betty Moore Foundation (GBMF4547 and GBMF9461; M.Z.H.). S.M.H.  acknowledges the support by the Ministry of Science and Technology, Taiwan (MOST) under Grant No. MOST 108-2112-M-110-013-MY3. C-C. L. acknowledges the Ministry of Science and Technology of Taiwan for financial support under contract No. MOST 110-2112-M-032-016-MY2. F.C.C. acknowledges  the support by the National Center for Theoretical Sciences and the Ministry of Science and Technology of Taiwan under grant nos. MOST-107-2628-M-110-001-MY3 and MOST-110-2112-M-110-013-MY3.  S.-Y.X. acknowledges the support of the Center for the Advancement of Topological Semimetals (CATS), an Energy Frontier Research Center (EFRC) funded by the US Department of Energy (DOE) Office of Science, through the Ames Laboratory under contract DE-AC0207CH11358 (fabrication and measurements), the STC Center for Integrated Quantum Materials (CIQM), National Science Foundation (NSF) award no. ECCS-2025158 (data analysis), and the Corning Fund for Faculty Development. H.L. acknowledges the support of the Ministry of Science and Technology (MOST) in Taiwan under grant number MOST 109-2112-M-001-014-MY3.  T.-R.C. was supported by the Young Scholar Fellowship Program from the Ministry of Science and Technology (MOST) in Taiwan, under a MOST grant for the Columbus Program MOST110-2636-M-006-016, the National Cheng Kung University, Taiwan, and National Center for Theoretical Sciences, Taiwan. Work at NCKU was supported by MOST, Taiwan, under grant MOST107-2627-E-006-001 and Higher Education Sprout Project, Ministry of Education to the Headquarters of University Advancement at NCKU.

\clearpage

\clearpage
\newpage
\section{Methods}

\subsection{CDW tight-binding model in real space}

Consider a cubic system with one atom per unit cell and lattice parameter $a$. We construct a two-band model with four Weyl points with breaking inversion symmetry and preserve time-reversal symmetry, 
\begin{equation}
\mathcal{H}(k)=A\sin (k_{x}a)\sin (k_{z}a)\sigma_{0} +\left[ \cos (k_{x}a)-\cos (k_{1}a)\right]  \sigma_{x} +\sin (k_{y}a)\sigma_{y} +\left[ 1+\cos (k_{z}a)-\cos (k_{y}a)\right]  \sigma_{z}.
\end{equation}
By taking Fourier transform of the lattice tight-binding model in momentum space, we get the hopping parameters as following,
\begin{eqnarray}%
t_{ij} &=&\frac{1}{N} \sum^{BZ}_{\bm{k} } e^{i\bm{k} \cdot \bm{r} }\mathcal{H}(\bm{k} ) \\
&=&A\left\{ \frac{1}{2i} \left[ \delta (\bm{r} +a\hat{x} )-\delta (\bm{r} -a\hat{x} )\right]  \frac{1}{2i} \left[ \delta (\bm{r} +a\hat{z} )-\delta (\bm{r} -a\hat{z} )\right]  \right\}  \sigma_{0}  \notag \\  
&&+\left\{ \frac{1}{2} \left[ \delta (\bm{r} +a\hat{x} )-\delta (\bm{r} -a\hat{x} )\right]  -\delta (\bm{r} )\cos k_{1}\right\}  \sigma_{x} \notag \\
&&+ \left\{ \frac{1}{2i} \left[ \delta (\bm{r} +a\hat{y} )-\delta (\bm{r} -a\hat{y} )\right]  \right\}  \sigma_{y}   \notag \\
&&+\left\{ \delta (\bm{r} )+\frac{1}{2} \left[ \delta (\bm{r} +a\hat{z} )+\delta (\bm{r} -a\hat{z} )\right]  -\frac{1}{2} \left[ \delta (\bm{r} +a\hat{y} )+\delta (\bm{r} -a\hat{y} )\right]  \right\}  \sigma_{z},   
\end{eqnarray}%
where $\bm{r}=\bm{r_i}-\bm{r_j}$ and $\bm{r_i}$, $\bm{r_j}$ are the displacements of lattice sites, and $\sigma$ is the Pauli matrix which describes the orbital degree of freedom on the atom.
Let $a=1$ and the two bands Hamiltonian without interaction term is 
\begin{eqnarray}%
H_0 &=& \sum_{i,j} c^{\dag }_{i}t_{ij}c_{j} + \mathrm{H.c.}\\
&=&\sum_{i} \frac{-A}{4} \left[ c^{\dag }_{i}\sigma_{0} c_{i+\hat{x} +\hat{z} }+c^{\dag }_{i}\sigma_{0} c_{i-\hat{x} -\hat{z} }-c^{\dag }_{i}\sigma_{0} c_{i+\hat{x} -\hat{z} }-c^{\dag }_{i}\sigma_{0} c_{i-\hat{x} +\hat{z} }\right]   \notag\\ 
&&+\sum_{i} \left[ -\cos k_{1}\, c^{\dag }_{i}\sigma_{x} c_{i}+\frac{1}{2} c^{\dag }_{i}\sigma_{x} c_{i+\hat{x} }-\frac{1}{2} c^{\dag }_{i}\sigma_{x} c_{i-\hat{x} }\right]   \notag\\
&&+\sum_{i} \frac{1}{2i} \left[ c^{\dag }_{i}\sigma_{y} c_{i+\hat{y} }-c^{\dag }_{i}\sigma_{y} c_{i-\hat{y} }\right]   \notag \\
&&+\sum_{i} \left[ c^{\dag }_{i}\sigma_{z} c_{i}+\frac{1}{2} c^{\dag }_{i}\sigma_{z} c_{i+\hat{z} }+\frac{1}{2} c^{\dag }_{i}\sigma_{z} c_{i-\hat{z} }-\frac{1}{2} c^{\dag }_{i}\sigma_{z} c_{i+\hat{y} }-\frac{1}{2} c^{\dag }_{i}\sigma_{z} c_{i-\hat{y} }\right]  + \mathrm{H.c.}
\end{eqnarray}%
where $c_{i}=(c_{i,1},\; c_{i,2})$ and $c_{i,1},\; c_{i,2}$ are the elecctron annihilation operators with the orbital (pseudo-spin) index $1,\,2$ on the atom at the site $\bm{r_i}$.
Then we consider the semi-1D CDW as Peierls dimerization in a double supercell along $x$ direction. We build a double cell supercell along the $\hat{x}$ direction and we denote the two atoms within the unit cell at $\bm{r'_i}$ as $c_i$ and $d_i$, where $c_{i}=(c_{i,1},\; c_{i,2})$, $d_{i}=(d_{i,1},\; d_{i,2})$ and $\bm{r'_i}$ as the displacements of super lattice sites. We can write the Hamiltonian in the supercell basis as 

\begin{eqnarray}
H_{0,SC} &=&\sum_{i} \frac{-A}{4} \left[  c^{\dag }_{i}\tau_{0} d_{i+\hat{z} }+c^{\dag }_{i}\tau_{0} d_{i-\hat{x} -\hat{z} }-c^{\dag }_{i}\tau_{0} d_{i-\hat{z} }-c^{\dag }_{i}\tau_{0} d_{i-\hat{x} +\hat{z} } \right. \nonumber \\
&\;& + \left. d^{\dag }_{i}\tau_{0} c_{i+\hat{x} +\hat{z} }+d^{\dag }_{i}\tau_{0} c_{i-\hat{z} }-d^{\dag }_{i}\tau_{0} c_{i+\hat{x} -\hat{z} }-d^{\dag }_{i}\tau_{0} c_{i+\hat{z} }\right]   \nonumber \\ 
&+&\sum_{i} \left[ -\cos k_{1}\, (c^{\dag }_{i}\sigma_{x} c_{i}+d^{\dag }_{i}\sigma_{x} d_{i})+\frac{1}{2} c^{\dag }_{i}\tau_{x} d_{i}-\frac{1}{2} c^{\dag }_{i}\tau_{x} d_{i-\hat{x} }+\frac{1}{2} d^{\dag }_{i}\tau_{x} c_{i+\hat{x} }-\frac{1}{2} d^{\dag }_{i}\tau_{x} c_{i}\right]    \nonumber \\
&+&\sum_{i} \frac{1}{2i} \left[ (c^{\dag }_{i}\sigma_{y} c_{i+\hat{y} }-c^{\dag }_{i}\sigma_{y} c_{i-\hat{y} })+(d^{\dag }_{i}\sigma_{y} d_{i+\hat{y} }-d^{\dag }_{i}\sigma_{y} d_{i-\hat{y} })\right]    \nonumber \\
&+&\sum_{i} \left[ c^{\dag }_{i}\sigma_{z} c_{i}+\frac{1}{2} c^{\dag }_{i}\sigma_{z} c_{i+\hat{z} }+\frac{1}{2} c^{\dag }_{i}\sigma_{z} c_{i-\hat{z} }-\frac{1}{2} c^{\dag }_{i}\sigma_{z} c_{i+\hat{y} }-\frac{1}{2} c^{\dag }_{i}\sigma_{z} c_{i-\hat{y} } \right.  \nonumber \\
&\;& + \left. d^{\dag }_{i}\sigma_{z} d_{i}+\frac{1}{2} d^{\dag }_{i}\sigma_{z} d_{i+\hat{z} }+\frac{1}{2} d^{\dag }_{i}\sigma_{z} d_{i-\hat{z} }-\frac{1}{2} d^{\dag }_{i}\sigma_{z} d_{i+\hat{y} }-\frac{1}{2} d^{\dag }_{i}\sigma_{z} d_{i-\hat{y} }\right]   \nonumber \\
&+& \mathrm{H.c.}
\end{eqnarray}%
where $\tau$ is the Pauli matrix describing the degree of freedom between two atoms.
Then, the Peierls dimerization can be realized by modifying the real space hopping strength between the two nearest neighbor atoms along the CDW direction in the supercell with a strength $\delta$, and the interaction terms can be expressed in the supercell basis as
\begin{eqnarray}
H_{int} &=&\sum_{i} \left[ \delta c^{\dag }_{i}\tau_{x} d_{i}-(-\delta )c^{\dag }_{i}\tau_{x} d_{i-\hat{x} }+(-\delta )d^{\dag }_{i}\tau_{x} c_{i+\hat{x} }-\delta d^{\dag }_{i}\tau_{x} c_{i}\right] + \mathrm{H.c.} .
\end{eqnarray}%
Then the full Hamiltonian becomes
\begin{eqnarray}
H_{CDW} &=& H_{0,SC}+ H_{int} \nonumber  \\
&=&\sum_{i} \frac{-A}{4} \left[  c^{\dag }_{i}\tau_{0} d_{i+\hat{z} }+c^{\dag }_{i}\tau_{0} d_{i-\hat{x} -\hat{z} }-c^{\dag }_{i}\tau_{0} d_{i-\hat{z} }-c^{\dag }_{i}\tau_{0} d_{i-\hat{x} +\hat{z} } \right. \nonumber \\
&\;& + \left. d^{\dag }_{i}\tau_{0} c_{i+\hat{x} +\hat{z} }+d^{\dag }_{i}\tau_{0} c_{i-\hat{z} }-d^{\dag }_{i}\tau_{0} c_{i+\hat{x} -\hat{z} }-d^{\dag }_{i}\tau_{0} c_{i+\hat{z} }\right]   \nonumber \\ 
&+&\sum_{i} \left[ -\cos k_{1}\, (c^{\dag }_{i}\sigma_{x} c_{i}+d^{\dag }_{i}\sigma_{x} d_{i})+(\frac{1}{2} +\delta )c^{\dag }_{i}\tau_{x} d_{i}-(\frac{1}{2} -\delta )c^{\dag }_{i}\tau_{x} d_{i-\hat{x} } \right. \nonumber \\
&\;& + \left. (\frac{1}{2} -\delta )d^{\dag }_{i}\tau_{x} c_{i+\hat{x} }-(\frac{1}{2} +\delta )d^{\dag }_{i}\tau_{x} c_{i}\right]      \nonumber \\
&+&\sum_{i} \frac{1}{2i} \left[ (c^{\dag }_{i}\sigma_{y} c_{i+\hat{y} }-c^{\dag }_{i}\sigma_{y} c_{i-\hat{y} })+(d^{\dag }_{i}\sigma_{y} d_{i+\hat{y} }-d^{\dag }_{i}\sigma_{y} d_{i-\hat{y} })\right]    \nonumber \\
&+&\sum_{i} \left[ c^{\dag }_{i}\sigma_{z} c_{i}+\frac{1}{2} c^{\dag }_{i}\sigma_{z} c_{i+\hat{z} }+\frac{1}{2} c^{\dag }_{i}\sigma_{z} c_{i-\hat{z} }-\frac{1}{2} c^{\dag }_{i}\sigma_{z} c_{i+\hat{y} }-\frac{1}{2} c^{\dag }_{i}\sigma_{z} c_{i-\hat{y} } \right.  \nonumber \\
&\;& + \left. d^{\dag }_{i}\sigma_{z} d_{i}+\frac{1}{2} d^{\dag }_{i}\sigma_{z} d_{i+\hat{z} }+\frac{1}{2} d^{\dag }_{i}\sigma_{z} d_{i-\hat{z} }-\frac{1}{2} d^{\dag }_{i}\sigma_{z} d_{i+\hat{y} }-\frac{1}{2} d^{\dag }_{i}\sigma_{z} d_{i-\hat{y} }\right]   \nonumber \\
&+& \mathrm{H.c.}
\end{eqnarray}%
It's worth noting that the $\delta \tau_x$ terms are in the off-diagonal blocks of the four orbitals basis, which reflects the nature of Weyl nodes interacting with each other through CDW.


\subsection{First-principles calulations}

First-principles calculations for (TaSe$_4$)$_2$I were performed using OpenMX code, where the generalized-gradient approximation, norm-conserving pseudopotentials, and optimized pseudoatomic basis functions were adopted \cite{OpenMX, DFT1, DFT2,DFT3,DFT4}. Three, two, two, and one optimized radial functions were allocated for the s, p, d, and f orbitals, respectively, for each Ta atom with a cut-off radius of 7 bohr, denoted as Ta7.0-s3p2d2f1. For the Se and I atoms, Se7.0-s3p3d2f1 and I7.0-s3p3d2f1 were adopted, respectively. A cutoff energy of 300 Ry was used for numerical integrations and for the solution of the Poisson equation. The first-principles calculations on Mo$_3$Al$_2$C were carried out using the Vienna Ab initio Simulation Package (VASP) with the projector augmented wave potentials \cite{DFT1}. The exchange-correlation functional was treated within the Perdew-Burke-Ernzerhof (PBE) generalized gradient approximations \cite{DFT3}.

\clearpage

\end{document}